# DISTRIBUTED KERNEL K-MEANS FOR LARGE SCALE CLUSTERING

Marco Jacopo Ferrarotti[1], Sergio Decherchi[1, 2] and Walter Rocchia[1]

[1] Istituto Italiano di Tecnologia, Genoa, Italy
sergio.decherchi@iit.it

[2] BiKi Technologies s.r.l, Genoa, Italy

## ABSTRACT

*Clustering samples according to an effective metric and/or vector space representation is a challenging unsupervised learning task with a wide spectrum of applications. Among several clustering algorithms, k-means and its kernelized version have still a wide audience because of their conceptual simplicity and efficacy. However, the systematic application of the kernelized version of k-means is hampered by its inherent square scaling in memory with the number of samples. In this contribution, we devise an approximate strategy to minimize the kernel k-means cost function in which the trade-off between accuracy and velocity is automatically ruled by the available system memory. Moreover, we define an ad-hoc parallelization scheme well suited for hybrid cpu-gpu state-of-the-art parallel architectures. We proved the effectiveness both of the approximation scheme and of the parallelization method on standard UCI datasets and on molecular dynamics (MD) data in the realm of computational chemistry. In this applicative domain, clustering can play a key role for both quantitively estimating kinetics rates via Markov State Models or to give qualitatively a human compatible summarization of the underlying chemical phenomenon under study. For these reasons, we selected it as a valuable real-world application scenario.*

## 1. INTRODUCTION

Grouping unlabelled data samples into meaningful groups is a challenging unsupervised Machine Learning (ML) problem with a wide spectrum of applications, ranging from image segmentation in computer vision to data modelling in computational chemistry [1]. Since 1957, when k-means was originally introduced, a plethora of different clustering algorithms arose without a clear all-around winner.

Among all the possibilities, k-means as originally proposed, is still widely adopted mainly because of its simplicity and the straightforward interpretation of its results. The applicability of such simple, yet powerful, algorithm however is limited by the fact that, by construction, it is able to correctly identify only linearly separable clusters and it does require an explicit feature space (i.e. a vector space where each sample has explicit coordinates).

To overcome both these limitations one can take advantage of the well-known kernel extension of k-means [2]. Computational complexity and memory occupancy are the major drawbacks of kernel k-means: the size of the kernel matrix to be stored together with the number of kernel function evaluations scales quadratically with the number of samples. This computational burden has historically limited the success of kernel k-means as an effective clustering technique. In fact, even though the potential of such approach has been theoretically demonstrated, few works in the literature [3] explore possibly more efficient approaches able to overcome the $O(N^2)$ computational cost.

We selected a real-world challenging application scenario, namely Molecular Dynamics (MD) simulations of biomolecules in the field of computational chemistry. Such atomistic simulations, obtained by numerical integration of the equations of motion, are a valuable tool in the study of biomolecular processes of paramount importance such as drug-target interaction [4]. MD simulations produce an enormous amount of data in the form of conformational frames (i.e. atoms positions at a given time step) that need to be processed and converted into humanly readable models to get mechanistic insights. Clustering can play a crucial role in this, as demonstrated by the success of recent works [1] and by the popularity of Markov state models [5]. We stress the fact that kernel k-means, without requiring an explicit feature space, is particularly suited for clustering MD conformational frames where roto-translational invariance is mandatory.

We introduce here an approximated kernel k-means algorithm together with an ad-hoc distribution strategy particularly suited for massively parallel hybrid CPU/GPU architectures. We reduce the number of kernel evaluations both via a mini-batch approach and an a priori sparse representation for the cluster centroids. As it will be clear, such twofold approximation is controlled via two straightforward parameters: the number of mini-batches $B$ and the sparsity degree of the centroid representation $s$. These two knobs allow to finely adapt the algorithm to the available computational resources to cope with virtually any sample size.

The rest of the paper is organized as follow: in section 2, we briefly review the standard kernel k-means [2] [6] algorithm. In section 3 our approximate approach is introduced together with a detailed description of the proposed distribution and acceleration strategy. Section 4 contains the assessment of both the approximation degree and the performances on standard ML datasets and a real case MD scenario. A discussion section together with conclusions complete the work.

## 2. KERNEL K-MEANS

Given a set $X$ of data samples $x_i \in \mathbb{R}^d, i \in [1, N]$, a non-linear transformation $\phi(x_i): \mathbb{R}^d \to \mathbb{R}^{d'}$ and said $C$ the number of clusters to be found, the kernel k-means algorithms finds a set $W$ of centroids $w_j \in \mathbb{R}^{d'}, j \in [1, C]$ in the transformed space, minimizing the following cost function:

$$\Omega(W) = \sum_{i=1}^{N} \sum_{j=1}^{C} \| \phi(x_i) - w_j \|^2 \delta(u_i, j) \qquad (1)$$

Where $u_i$ is the index of the closest prototype (i.e. the predicted label for sample $i$-th) obtained as:

$$u_i = \underset{j}{argmin} \| \phi(x_i) - w_j \|^2 \qquad (2)$$

and $\delta(u_i, j)$ is the usual Kronecker delta.

A Gradient Descent (GD) procedure can be used in order to locally minimize the non-convex cost $\Omega(W)$ starting from an initial set of cluster prototypes $W_0 = \{w_{j,0}\}$ so that at the $t$-th iteration we have:

$$w_{j,t} = \frac{1}{|w_{j,t}|} \sum_{i=1}^{N} \phi(x_i) \delta(u_{i,t}, j) \qquad (3)$$

where the $j$-th cluster cardinality is indicated as $|w_j| = \sum_{i=1}^{N} \delta(u_i, j)$.

A self-consistent update equation can be derived substituting Eq.3 into Eq.1:

$$u_{i,t+1} = \underset{j}{\operatorname{argmin}}\{\frac{1}{|w_{j,t}|^2}\sum_{m,n} K_{m,n}\delta(u_{m,t},j)\delta(u_{n,t},j) - \frac{2}{|w_{j,t}|}\sum_m K_{i,m}\delta(u_{m,t},j)\}$$
$$= \underset{j}{\operatorname{argmin}} g_{j,t} - 2f_{(i,j),t} \qquad (4)$$

Where the inner product in the transformed space $<\phi(x_m),\phi(x_n)>$ was replaced with a generic Mercer kernel $K(x_m, x_n) = K_{m,n}$ and where we introduced the cluster compactness and cluster average similarity respectively defined as:

$$g_j = \frac{1}{|w_j|^2}\sum_{m,n} K_{m,n}\delta(u_m,j)\delta(u_n,j) \qquad (5)$$

$$f_{i,j} = \frac{1}{|w_j|}\sum_m K_{i,m}\delta(u_m,j) \qquad (6)$$

It is therefore clear that the knowledge of the kernel matrix is sufficient to update the set of predicted labels up to convergence. Since an explicit form for $\phi(x)$ is not known in general, a medoid approximation can then be used in order to obtain an approximated estimate of the cluster prototypes:

$$\phi^{-1}(w_j) \approx m_j = \underset{x_l \in X}{\operatorname{argmin}} \| \phi(x_l) - w_j \|^2$$
$$= \underset{x_l \in X}{\operatorname{argmin}} K_{l,l} - 2\frac{1}{|w_j|}\sum_i K_{i,l}\delta(u_i,j) \qquad (7)$$
$$= \underset{x_l \in X}{\operatorname{argmin}} K_{l,l} - 2f_{i,j}$$

As shown in [7], for the linear case, the kind of iterative algorithm described by Eq.4 almost surely converge to a local minimum, eventually reaching the stopping condition $u_{i,t+1} = u_{i,t}, \quad \forall i \in [1,N]$.

We conclude this section with a final remark on the cluster compactness and the cluster average similarity (i.e. Eq.5-6). Indeed a kernel k-means reformulation in term of such quantities was originally proposed by Zhang and Rudnicky [6] in order to reduce the memory footprint of the kernel matrix allowing caching on disk. As we are going to show in the next section, the same formalism can be effectively used to design an efficient distribution strategy.

## 3. DISTRIBUTED MINI-BATCH KERNEL K-MEANS

We present in this section our contribution: a novel approximation for the kernel k-means algorithm together with an ad-hoc distribution and acceleration strategy well suited for nowadays heterogenous High Performance Computing (HPC) facilities.

***Remark about the notation used:*** in the following a superscript eventually identifies a specific mini-batch quantity, when no superscript is used the quantity has to be intended as a global quantity. As an example $w_j^i$ represents the $j$-th cluster prototype for the $i$-th mini-batch whereas $w_j$ is the $j$-th global cluster prototype obtained combining the partial results of every mini-batches.

### 3.1. The Mini-batch Kernel K-Means

Our primary approach to reduce the $O(N^2)$ complexity coming from the kernel matrix evaluation consists of splitting the dataset in disjoint mini-batches that are processed one after the other. The procedure can be summarized by these steps:

1. Fetch one mini-batch at a time until all data is consumed.

2. Perform kernel k-means clustering on one minibatch and collect results with a proper initialization technique.

3. Merge together current minibatch results to global results with a proper strategy and go to step 1.

Fig.1(a) shows a pictorial description of such algorithm highlighting its hierarchical structure. The entire procedure is detailed in the subsequent paragraphs.

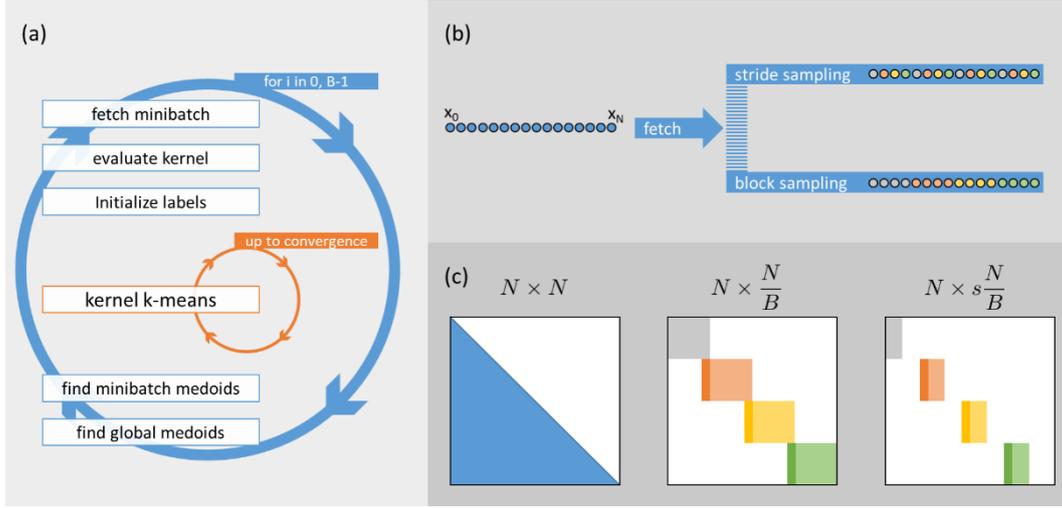

Fig. 1 (a) Pictorial description of the algorithm. (b) Visualization of two possible sampling strategies to divide the dataset into mini-batches. (c) From left to right we visualize the effect of the two fold approximation proposed on the number of kernel matrix elements that need to be evaluated.

*Mini-batch fetching:* The first sensible choice to be made, regards the way in which the dataset is divided in $B$ disjoint mini-batches of size $N^i, \forall i \in [0, B-1]$. Without loss of generality we will consider in the following $N^i = \frac{N}{B} \forall i \in [0, B-1]$. A variety of possibilities arise, we present here two common reasonable sampling strategies.

A stride sampling strategy can be used when the entire dataset is known beforehand and one wants to minimize the correlations among samples within the same mini-batch i.e. $X^i = \{x_{i+jB}\}, j \in [0, \frac{N}{B} - 1]$.

A block sampling strategy can be used instead to process a data stream in order to start the clustering procedure as soon as the first $N^0$ samples are received i.e. $X^i = \{x_{i\frac{N}{B}+j}\}, j \in [0, \frac{N}{B} - 1]$.

For the sake of clarity the two different sampling strategies presented are visualized in Fig.1(b).

*Kernel evaluation and mini-batch initialization:* Once a mini-batch is fetched, it is straightforward to evaluate the mini-batch kernel matrix $K^i$ with a computational cost of $O(\frac{N^2}{B^2})$. Let us now discuss how it is possible to initialize the $i$-th mini-batch labels. We distinguish two cases:

$i = 0$: during the first mini-batch the global cluster medoids have to be selected randomly or by means of some rational. We propose here to use a kernelized version of the popular k-means++ initialization scheme, where the medoids are picked at random with a distribution that maximize the distance among them. The interested reader can read the work in [8] where such initialization scheme is discussed in detail for the linear case.

$i \neq 0$: Starting from the second mini-batch the global cluster medoids $M = \{m_j \approx \phi^{-1}(w_j)\}$ obtained at the end of the previous iterations are used for the initialization. Simply applying Eq.2 we have:

$$u_l^i = \arg\min_j [K(x_l^i, x_l^i) - 2K(x_l^i, \tilde{x}_j)] \tag{8}$$

Such initialization step automatically allows to keep track of the clusters across different mini-batches. Indeed the global $j$-th medoid obtained at the end of the $(i-1)$-th iteration is used as initialization for the same $j$-th cluster of the $i$-th mini-batch. This avoids ambiguity also when the partial mini-batch result has to be merged with the global one. The mini-batch medoid $m_j^i$ will be combined with the global centroid $m_j$ having the same index $j$.

It should be understood that in order to evaluate the second term of Eq.8 one has to perform additional computations. One has to compute the kernel function for all the pairs $(x_l^i, m_j)$ where $x_l^i$ belongs to the $i$-th mini-batch and $m_j$ its a global medoid coming from the $(i-1)$-th mini-batch. Thus, the initialization phase of each mini-batch requires the evaluation of the corresponding auxiliary kernel matrix $\tilde{K}^i$ of size $\frac{N}{B} \times C$.

*Mini-batch inner GD loop:* Given a mini-batch kernel matrix $K^i$ and an initial set of labels $U_0^i$, equations Eq.2-5 are used to perform a GD optimization of the reduced cost function:

$$\Omega(W^i) = \sum_{x_j \in X^i} \sum_{l=1}^{C} \| \phi(x_j) - w_l^i \|^2 \, \delta(u_j^i, l) \tag{9}$$

A final set of labels $U^i$ is obtained as a result of such optimization procedure. It is worth stressing the fact that at this point the set of mini-batch cluster prototypes is not known in terms of explicit coordinates, but just in term of membership. As a solution, we propose the medoid approximation introduced in section 2. Using equation Eq.7, we set the cluster prototypes as:

$$w_j^i \leftarrow \phi(m_j^i): \quad m_j^i = \arg\min_{x_l \in X^i} \| \phi(x_l) - w_j^i \|^2 \tag{10}$$

More sophisticated approaches based, for instance, on a sparse representation of cluster centres are possible (e.g. see [9]). However, the inherent additional computational cost and the satisfactory results already obtained by means of the simple medoid approximation discouraged us to further investigate this possibility.

*Full batch cluster centres update:* We discuss now on how to merge the medoids $M^i$ of the $i$-th mini-batch together with the global medoid set $M$. Let $\{w_j = \phi(m_j)\}$ be the global medoids at the $(i-1)$-th iteration of the outer loop and let $\{w_j^i = \phi(m_j^i)\}$ be the cluster centres for the current $i$-th mini-batch. We propose to obtain the resulting global cluster prototypes as a convex combination of the two:

$$w_j \leftarrow (1-\alpha)\phi(m_j) + \alpha\phi(m_j^i) \tag{11}$$

Practically, since Eq.11 cannot be evaluated directly, we introduce a second medoid approximation as already done in the previous paragraph, so that:

$$w_j \leftarrow \phi(m_j): \quad m_j \leftarrow \arg\min_{x_l \in X^i} \| \phi(x_l) - (1-\alpha)\phi(m_j) - \alpha\phi(m_j^i) \|^2 \quad (12)$$

The choice of this convex combination stems from a simple but important observation; in order to choose the coefficient $\alpha$ let us consider the updating equation for the global cluster center $w_j$ at the second iteration of the algorithm, when the first two mini-batches are merged in a single one (assuming this is the complete dataset):

$$\begin{aligned}
w_j &= \frac{1}{|w_j^0|+|w_j^1|}\sum_{x_i \in X^0 \cup X^1} \phi(x_i)\delta(u_i, j) \\
&= \frac{|w_j^0|}{|w_j^0|+|w_j^1|}\frac{1}{|w_j^0|}\sum_{x_i \in X^0} \phi(x_i)\delta(u_i, j) + \frac{|w_j^1|}{|w_j^0|+|w_j^1|}\frac{1}{|w_j^1|}\sum_{x_i \in X^1} \phi(x_i)\delta(u_i, j) \quad (13) \\
&= \frac{|w_j^0|}{|w_j^0|+|w_j^1|}w_j^0 + (1 - \frac{|w_j^0|}{|w_j^0|+|w_j^1|})w_j^1
\end{aligned}$$

We therefore set $\alpha = \frac{|w_j^i|}{|w_j^i|+|w_j|}$ so that, if each mini-batch is labelled correctly at the end of the GD minimization, we retrieve the correct result (i.e. same cluster medoids as for full batch kernel k-means).

***Empty clusters:*** We close this subsection with a remark about empty-clusters. It is not guaranteed that along inner loop iterations there will be at least one data sample per cluster. This is a well-known k-means issue and several strategies to deal with such empty-clusters problem are possible e.g. randomly pick a new cluster prototype or reducing $C$. Here we propose the following: if a given cluster $j$ is found to be empty at the end of the $i$-th mini-batch iteration then its global prototype will not be updated. It is worth noting that this kind of strategy is naturally embedded in the definition of $\alpha$ since for $|w_j^i| = 0$ we have $\alpha = 0$ and Eq.11 guarantee the correct behaviour.

### 3.2. Approximate Mini-batch Kernel K-Means

In the previous paragraph we introduced a simple yet powerful mini-batch approximation which allowed us to reduce the number of kernel evaluations down to $N\frac{N}{B}$. Here, we show how we can further reduce the complexity of the algorithm by means of an a priori sparse representation of the cluster centroids. This approach was first introduced by Chitta et al. and relies on the simple observation that the full kernel matrix is required at each iteration of the kernel k-means algorithm because the clusters centres are represented as a linear combination of the entire dataset. However, the number of kernel elements to be evaluated can be drastically reduced if one restricts the cluster centres to a smaller sub space spanned by a small number of landmarks i.e. data samples randomly extracted from the dataset. A complete review of such approximation technique is out of the scope of this work, the interested reader can refer to [3] for further details.

We limit ourselves to illustrate here how we can reformulate the same idea within our algorithm. In order to do so we simply need to restrict the summation in Eq.3 on the subset $i: x_i \in L$ where $L = \{l_0, \ldots, l_{|L|}\}$ is a set of landmarks uniformly sampled from the mini-batch.

$$w_j = \frac{1}{|w_j|}\sum_{i \in L} \phi(x_i)\delta(u_i, j), j \in [1, C] \quad (14)$$

The self-consistent update equation for the minibatch labels will be:

$$u_i^{t+1} = \operatorname*{argmin}_j [\hat{g}(w_j^t) - 2\hat{f}(x_i, w_j^t)] \tag{15}$$

where $\hat{g}(w_j)$ and $\hat{f}(x_i, w_j)$ are the approximate mini-batch clusters compactness and mini-batch clusters similarity

$$\hat{g}(w_j) = \frac{1}{|w_j^t|^2} \sum_{m,n \in L} K_{m,n} \delta(u_m^t, w_j^t) \delta(u_n^t, w_j^t) \tag{16}$$

$$\hat{f}(x_i, w_j) = \frac{1}{|w_j^t|} \sum_{m \in L} K_{i,m} \delta(u_m^t, w_j^t) \tag{17}$$

It should be clear from Eq.16 and Eq.17 that the number of kernel evaluations needed to run such approximated algorithm is now $N|L| = sN\frac{N}{B}$, where the key parameter $s$ is the fraction of data used for the cluster centres representation in each mini-batch defined as:

$$s = \frac{|L|}{N} B \tag{18}$$

In Fig.1(c) the reader can visualize the effects that $B$ and $s$ have on the number of kernel elements that needed to be evaluated in order to iterate the proposed algorithm. As already stated in the introduction, these two parameters act like knobs that control the degree of approximation of the procedure with respect to standard kernel k-means. Later, we will discuss on how to pick proper values for these parameters according to the available computational resources.

### 3.3. Heterogeneous HPC implementation strategy

We discuss here how the nature of the previously introduced algorithm is particularly suited to be implemented on both distributed systems and heterogeneous architectures where an accelerator (e.g. general-purpose GPU) is paired to a CPU.

As already discussed in section 2, the whole iterative procedure to update the set of predicted labels minimizing the kernel k-means cost function can be expressed in terms of the average cluster similarity $f_i, j, \forall i \in 0, \ldots, \frac{N}{B}, j \in 0, \ldots, C-1$ and the cluster compactness $g_j \forall j \in 0, \ldots, C-1$. Both quantities can be expressed as partial summations of kernel matrix elements, where the elements to be summed are selected according to the labels via $\delta(u_i, j)$. From Eq.6 it should be clear that the summation to compute the $i$-th row of $f$ runs just over the $i$-th row of $K$, this naturally suggest us a row wise distribution strategy. Considering a system with $P$ nodes, the workload is divided so that each node $p$ accounts for the computation of $K_{i,j}$ and $f_{i,l} \forall j \in [0, \frac{N}{B}), i \in [p\frac{N}{BP}, (p+1)\frac{N}{BP}), l \in [0, C)$.

The full data distribution scheme is presented in Fig.2(a) and the resulting algorithm is detailed via pseudo code in Alg.1. The advantage of such approach mainly consists in the reduced communication overhead. Indeed, for each iteration of the inner loop two communication steps are sufficient, involving a reduction of the cluster compactness $g$ together with a gathering step for the updated labels $U$. The kernel matrix elements always reside locally to the node and they never go through the network.

The memory footprint can be easily computed and amounts to $Q(\frac{N}{BP}(\frac{N}{B} + C) + \frac{N}{B} + 2C)$ where $Q$ is the size of variables expressed in Bytes, this is a central quantity because in a real application scenario once fixed the computational resources i.e. amount of memory available per processor $R$ and the number of processors $P$, it allows us to compute the minimum number of mini-batches that can be used in order to process the entire dataset:

$$B_{min} = \frac{\frac{2N}{P}}{-(\frac{C}{P}+1)+\sqrt{(\frac{C}{P}+1)^2 - 8\frac{C}{P}+\frac{R}{Q}}} \quad (19)$$

An upper bound for the message size per node can also be easily given by $Q(\frac{N}{BP} + 2C)$. This however represents a worst-case scenario, where the entire set of labels $U$ are communicated at each step, instead of communicating just the ones that were actually updated.

The computational complexity of the proposed implementation grows as $O(\frac{N^2}{B^2 P})$ and it is dominated by the kernel matrix evaluation step. It is worth stressing the fact that we decided not to exploit any kernel matrix symmetry because that would have resulted in the impossibility of pursuing our row-wise data distribution scheme and additionally it would have hindered the possibility of using non symmetric similarity functions. Moreover, exploiting the kernel matrix symmetry would have resulted in a non trivial addressing scheme, unsuitable for the limited memory addressing capabilities of accelerators such as general purpose GPUs; this increased memory footprint is largely compensated by the approximation strategy in performance terms.

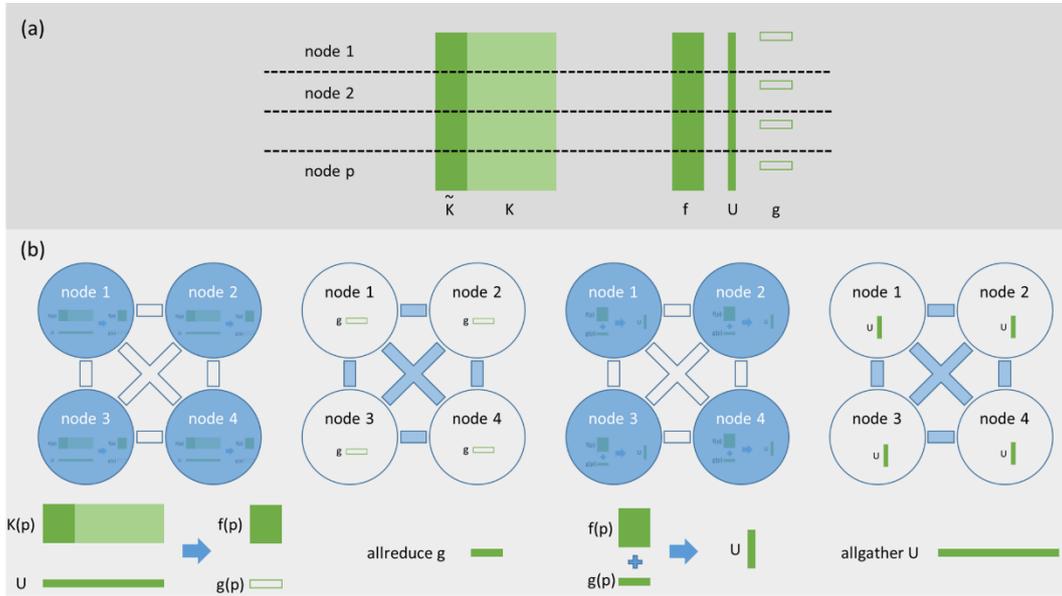

Fig. 2 (a) Distribution scheme for the principal quantities needed to complete an inner loop iteration. Each node holds a set of entire rows for $\widetilde{K}, K, f$ and $U$. Each node holds a local copy of $g$. (b) From left to right the main steps of an inner loop iteration are illustrated. At first, each node is computing its portion of $f$ together with a partial $g$ starting from its $K$ rows and $U$. Then, the global $g$ is retrieved with an all-to-all reduction step. In the third stage each node uses that information together with its portion of $f$ to compute its slice of $U$. As a final step an all-to-all gathering step spread the updated labels across the network.

| | | |
|---|---|---|
| **input:** | dataset $X$; number of clusters $C$; number of mini-batches $B$ | |
| **output:** | medoids $M$ | |
| 1 | **for** $i \leftarrow 1$ **to** $B$ **do** | |
| 2 |     $X^i \leftarrow$ samples fetched from $X \setminus X^{j<i}$ | |
| 3 |     $K^i(p) \leftarrow$ precompute kernel matrix | |
| 4 |     **if** $i == 0$ | |
| 5 |         $M^0 \leftarrow$ initialize according to kernel k-means++ | |
| 6 |     **end** | |
| 7 |     $U^i(p) \leftarrow$ assigned according nearest neighbor medoid | |
| 8 |     $t \leftarrow 0$ | |
| 9 |     **while** $U_t^i \mathrel{!=} U_{t+1}^i$ | |
| 10 |         **allgather** $U_t^i$ | sync |
| 11 |         $g^i(p) \leftarrow$ compute according to Eq.5 | |
| 12 |         $f^i(p) \leftarrow$ compute according to Eq.6 | |
| 13 |         **allreduce sum** $g^i$ | sync |
| 14 |         $U_{t+1}^i(p) \leftarrow$ assign accoding to Eq.4 | |
| 15 |         $t \leftarrow t+1$ | |
| 16 |     **end** | |
| 17 |     $M^i(p) \leftarrow$ medoid approximation according to Eq.10 | |
| 18 |     **allreduce min** $M^i$ | sync |
| 19 |     $M(p) \leftarrow (1-\alpha)M + \alpha M^i(p)$ | |
| 20 |     **allreduce min** $M$ | sync |
| 21 | **end** | |

Alg. 1 Distributed mini-batch kernel k-means pseudocode for node $p$.

Starting from this observation we discuss now how the mini-batch structure of the algorithm can be exploited in order to design an effective acceleration strategy. In the following we will consider an offload acceleration model where host processor and target device have separate memory address spaces and communicate via a bus with limited bandwidth (e.g. PCIe) with respect to the processor-memory standard bus.

The evaluation of a large kernel matrix perfectly fits the massively parallel architecture of nowadays accelerators therefore it seems a reasonable choice to offload that portion of the computation. One of the key element for an efficient acceleration scheme however is the overlapping between the host and the target workload [10], so that a simple strategy where the CPU and the accelerator are alternatively in idle waiting for each other is not desirable.

Each iteration $i$-th of the outer loop depends on the previous one, namely the $(i-1)$-th, in order to initialize the set of labels $U^i$. This is what prevents the algorithm to be trivially parallel forcing to run just one mini-batch per time. However, if one considers the first two steps of each outer loop iteration i.e. mini-batch fetch $X^i$ and kernel matrix evaluation $K^i$ it is clear that they can be performed independently for each $i$. We exploit this feature, instructing the target device to compute the kernel matrix $K^{(i+1)}$ while the host processor executes the inner loop of the algorithm on the $i$-th mini-batch.

The offload procedure is detailed in Fig.3; the overall performance gain heavily depends on the accelerator side implementation of the kernel matrix evaluation which goes outside the scope of the proposed paper.

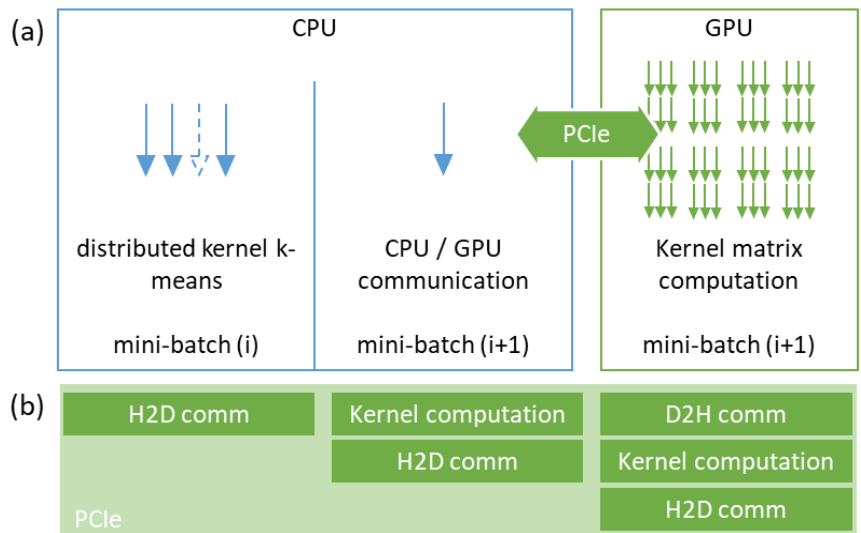

Fig. 3 (a) Pictorial description of the proposed acceleration scheme. The diagram is divided in two parts: a host processor side on the left, and a target device side on the right. We illustrate how multiple CPU threads can be used to overlap host and device workload. A CPU thread is bound to the device, it is responsible for data fetching from disk, for host-device data transfer and for device control. It instructs the device to compute the kernel matrix elements needed by the next $(i+1)$-th iteration of the outer loop. All the other available threads cooperate and are responsible for the current $i$-th iteration consuming the kernel matrix elements provided by the accelerator. In this sense device and host work in a producer-consumer pattern. (b) We detailed how a 3-stage pipeline can be used on the device in order to overlap the kernel computation with the host to device (H2D) and device to host (D2H) slow communications needed to transfer the dataset on the device and the kernel matrix back to host.

## 4. EXPERIMENTS

We implemented the proposed method and we present here some tests against standard datasets in the ML field as well as against a 2D toy dataset in order to better asses both performances and the degree of approximation. Moreover, we present an applicative scenario in the Computational Chemistry realm.

*2D Toy:* Synthetic dataset containing 4 clusters of 10000 elements in a 2D feature space. Each cluster is generated by sampling a Gaussian distribution with center and width carefully selected in order to facilitate its visualization i.e. ($\sigma$=[0.2,0,2],$\mu$=[0.25,0.75]), ($\sigma$=[0.2,0,2],$\mu$=[0.75,0.75]) and ($\sigma$=[0.2,0,2],$\mu$=[0.25,0.75]) .

*MNIST:* dataset of handwritten digits [11]. It is composed by a training set of 60000 samples and a test set of 30000 samples. 784-dimensional feature space with integer features.

**RCV1:** Reuters Corpus Volume I is a collection of manually labelled documents used as standard benchmark for classification in the domain of multilingual text categorization [12]. It is composed of 23149 training samples and 781265 test samples. Among the various formats available we used here its expression as normalized log TF-IDF (i.e. logarithmic term frequency-inverse document frequency) vectors in a sparse 47236-dimensional feature space. As already proposed in [13] we pre-processed the dataset removing samples with multiple labels and categories with less than 500 samples. After doing this we obtained a dataset of 193844 samples all coming from the test samples which we arbitrarily divided in 188000 training samples and 5844 test samples to maintain the original ratio. Moreover, to deal with the sparsity of the feature space we performed a dimensionality reduction step via random projection on a dense 256-dimensional space.

***Noisy MNIST:*** generated by starting from MNIST and adding uniform noise on 20% of the features. Each sample in the training set is perturbed 20 times in order to obtain a final dataset of 1200000 samples in a 784-dimensional normalized feature space.

***MD trajectory:*** As previously anticipated, we used Molecular Dynamics as an appealing clustering scenario in which to leverage the features of the proposed algorithm. Microsecond-long trajectories of the binding mechanism of a drug, specifically a transition state analogue named DADMe-immucillin-H, to the Purine Nucleoside Phosphorylase (PNP) enzyme were employed [14]. Those long trajectories well represent a good and relatively novel application domain for clustering and machine learning in general.

When possible, we compared the clustering labels coming from the proposed procedure with the training labels. We will consider mainly two standard quality measures:

***Clustering accuracy:*** Let $u_i$ be the set of labels obtained as a clustering result and let $y_i$ be the set of the actual classes given as training or test. The clustering accuracy is defined as $\mu(y,u) = \sum_{i=0}^{N-1} \frac{\delta(\psi(y_i),u_i)}{N}$. Where $\psi(u_i)$ is a mapping function which maps each clustering label to an actual training or test class. We propose here the use of a simple majority voting scheme to obtain such a mapping.

***Normalized Mutual Information:*** Let now be $n_i = \sum_{j=0}^{N-1} \delta(u_i,j)$ , $m_i = \sum_{j=0}^{N-1} \delta(y_i,j)$ and $o_{i,j} = \sum_{k=0}^{N-1} \delta(u_k,i)\delta(y_k,j)$ the normalized mutual information is a quality measure defined as

$$NMI(y,u) = \frac{\sum_{i,j} o_{i,j} \log(\frac{No_{i,j}}{n_i m_j})}{(\sum_i n_i \log(\frac{n_i}{N}))(\sum_i m_i \log(\frac{m_i}{N}))}$$

We tested our implementation on a variety of different platforms in order to better describe the versatility and the potential impact of the proposed algorithm:

***IBM-BG/Q - Cineca/FERMI:*** Cluster of 10240 computing nodes equipped with two octacore IBM PowerA2, 1.6 GHz processors each, for a total of 163840 cores. The available memory amounts to 16 GB / core and the internal network features a 5D toroidal topology.

***IBM NeXtScale - Cineca/GALILEO:*** Cluster of 516 computing nodes equipped with two octacore Intel Haswell 2.40 GHz processors for a total of 8256 cores. The available memory amounts to 8 GB / core and the internal network features Infiniband with 4x QDR switches.

***State-of-the-art Workstation:*** Modern desktop machine equipped with two Intel E-6500 esacore processors and 64 GByte of memory.

**4.1. Explanatory 2D toy model**

As a first step to assess the proposed clustering algorithm we consider the 2D Toy dataset. We aim at better illustrating and helping the visualization of the evolution of the cluster centres along with the iterations of the outer loop. Incidentally, we want to highlight the consequences of a poor sampling strategy (concept-drift) and to give a rationale for understanding its quality.

In figure 4(a)-(b) the evolution of the cluster centres is followed for two different sampling strategies i.e. (a) stride sampling and (b) block sampling. Even though the final set of labels is the same for such simple dataset it should be clear that the stride sampling strategy is superior in representing the structure of the dataset within each mini-batch. The underlying question is how could one assess the quality of the sampling strategy in a real case scenario where direct visualization is not possible. In Fig.4(c) we try to answer by looking at the behaviour of the cluster center displacement. We can comment that if such quantity is constantly small with respect to the average cluster size, the mini-batches can be regarded as good representative of the entire dataset structure. In contrast, high values or spikes in the same quantity may reflect a poor sampling strategy.

Observing Fig.4(d) we note that the inner loop of the proposed algorithm, i.e. the minimization of the partial cost $\Omega(W^i)$, does indeed help in minimizing the global objective function $\Omega(W)$.

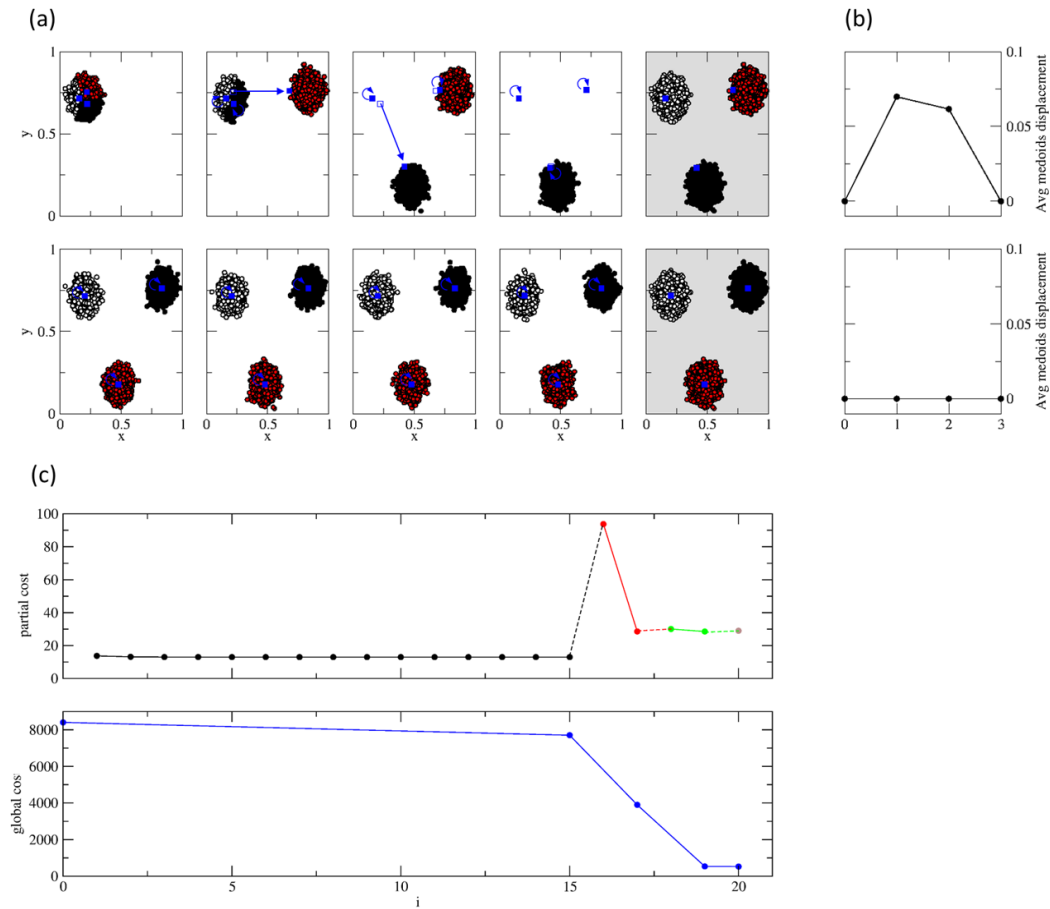

Fig. 4 (a-top row) From left to right the evolution of the cluster centres across different iterations of the outer loop in the case of a poorly designed block sampling strategy. (a-bottom row) From left to right the evolution of the cluster centres across different iterations of the outer loop in the case of a proper stride sampling strategy where each mini-batch correctly captures the underlying structure of data. (b) Average cluster centres displacement vs outer loop iterations for the two different sampling strategies illustrated in (a), we propose this as a control observable to assess the quality of the sampling when direct visualization is not feasible. (c-top panel) Partial cost function $\Omega(W^i), \forall i \in [0, B = 3]$ vs number of iterations, different colors represent different mini-batches. (c-bottom panel) Global cost function $\Omega(W)$ vs number of iterations. It is worth noting how the inner loop iterations within each mini-batch help to bring down the global cost function.

### 4.2. Assessing the degree of approximation

We consider now the MNIST dataset in order to assess the degree of approximation introduced by the mini-batch approach and by the a priori sparse representation of the cluster centres. We ran our algorithm on the 60000 training samples of MNIST with $B = [1,2,4,8]$, $s \in [0.025, 1.0]$ and we monitored the resulting clustering centres against the 10000 test samples in order to compute the clustering accuracy $\mu$. Results as well as execution times are presented in Fig.5. We observe that the algorithm is generally robust across a wide range of the two parameters. The clustering accuracy slightly decreases when the number of mini-batches increase and once $B$ is fixed it decreases almost monotonically with $s$ dropping to low values when $s < 0.2$. As expected, this suggests us to position ourselves to the top-left part of the graph i.e. few mini-batches and $s \approx 1$.

Both $B$ and $s$ are trade-off parameters that have to be fixed. The strategy we suggest here is to fix a desired execution time on a given compute architecture. The available memory for the execution can lead to a first value for $B$ using Eq.19. As a starting point, the value of $s$ can be fixed at its maximum. This set of parameters i.e. $(B_{min}, 1.0)$. should be optimal for the computational resources available i.e. minimum number of mini-batches without sparse representation of the cluster centroids. One can evaluate the expected execution time for the algorithm running it on a single mini-batch, if the expected execution time does not match the initial requirements then one can first slowly decrease $s$ and, if this is not sufficient (i.e. expected execution time too high for $s < 0.2$), then increase the number of mini-batches. The approximation degree introduced can be self consistently checked using a single mini-batch and taking as reference the results obtained for the optimal set of parameters $(B_{min}, 1.0)$.

This rationale should guide the user to finely tune the trade-off parameters also on a very large dataset.

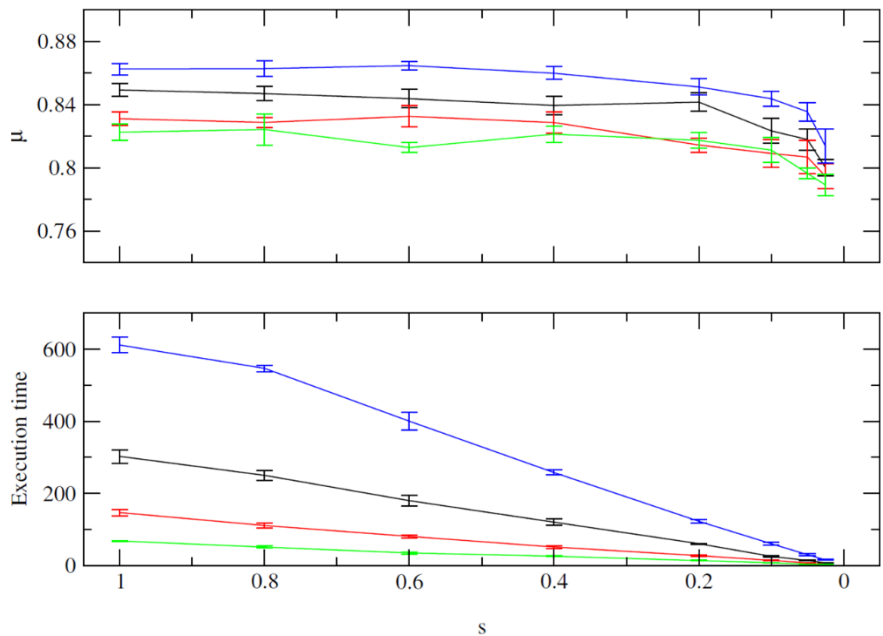

Fig. 5 (top panel) Cluster accuracy vs $s$. (bottom panel) Execution time vs $s$. Clustering performed on 60000 MNIST training samples evaluated against the 10000 provided test samples. Different colors represent different values of $B = [1,2,4,8]$. As described in the main text this graph can help understand how to perform model selection for the set of newly introduced parameters $(B, s)$ picking a target execution time and looking at the clustering accuracy for the compatible sets of parameters.

### 4.3. Scaling behaviour

We aim here at assessing the quality of the ad-hoc distribution strategy that we proposed in the previous section. In order to do so we tested our algorithm both on the IBM BG/Q and on the IBM NeXtScale machines above described, against the standard MNIST dataset.

We decided to set $B = 1$ in order to run the code in single batch mode since, as already explained, our distribution strategy does not involve the outer loop of the proposed method i.e. increasing the number of mini-batches would have only added a multiplicative constant to the execution time equal to $B$.

In Fig.6 the strong scaling plot for both machines is showed, the algorithm exhibits near to perfect scaling for a wide range of $P$ i.e. $16 \rightarrow 1024$ on IBM BG/Q and $16 \rightarrow 256$ on IBM NeXtScale. The discrepancy from the ideal behaviour outside this range can be ascribed to the portion of code intrinsically serial (e.g. fetching and initialization phases) which becomes a prominent cost as described by Amdahl's law.

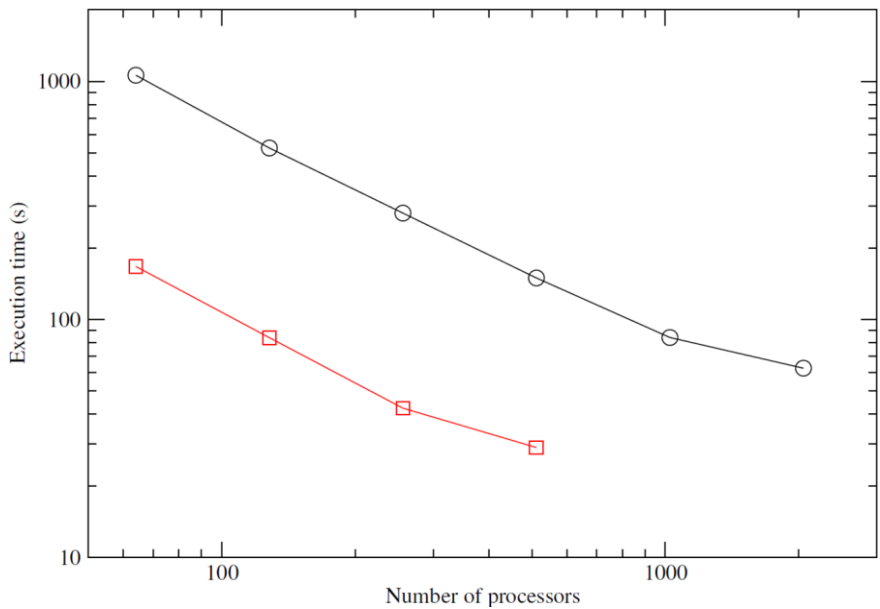

Fig. 6 Execution time vs $P$ for two different distributed architectures. IBM BG/Q in black/circles and IBM NeXtScale in red/squares.

### 4.4. Standard datasets analysis

We present here the tests we performed on a state-of-the-art workstation over standard datasets coming from the Machine Learning community. We show how even a large dataset with up to $10^6$ elements in 784 dimensions can be processed via a kernel approach on a desktop machine in a reasonable amount of time. The considered datasets are MNIST (60000 samples in 784 dimensions), RCV1 (188000 samples in 256 dimensions) and noisy MNIST (1000000 samples in 784 dimensions). The results are collected respectively in Tab.1-3.

For all the experiments, we used the strided sampling technique, set $s = 1$, selected the number of clusters automatically via the elbow criterion and set $\sigma = 4d_{max}$ in order to mimic a linear kernel behaviour. As a baseline comparison for the clustering accuracy and the normalized mutual information we used a standard python implementation of k-means from the scikit-learn package [15]. Results coming from RCV1 are also compared with that appearing in the literature [13].

Tab. 1 MNIST results and timings for different $B$ values

| B | Clustering accuracy | NMI | Execution time |
|---|---|---|---|
| Baseline | 84.5 ± 0.62 | 0.693 ± 0.012 | – |
| 1 | 86.47 ± 0.37 | 0.737 ± 0.006 | 655.23 ± 82.92 |
| 4 | 82.63 ± 0.91 | 0.680 ± 0.011 | 133.63 ± 4.40 |
| 16 | 81.45 ± 0.653 | 0.670 ± 0.010 | 32.17 ± 2.48 |
| 64 | 78.39 ± 0.95 | 0.626 ± 0.015 | 9.51 ± 0.58 |

Tab. 2 RCV1 results and timings for different $B$ values

| B | Clustering accuracy | NMI | Execution time |
|---|---|---|---|
| Literature | 16.59 ± 0.62 | 0.2737 ± 0.0063 | – |
| Baseline | 15.16 ± 0.81 | 0.091 ± 0.0052 | – |
| 4 | 17.41 ± 0.83 | 0.147 ± 0.006 | 797.65 ± 53.48 |
| 16 | 16.52 ± 0.74 | 0.145 ± 0.001 | 170.96 ± 4.94 |
| 64 | 16.15 ± 0.60 | 0.132 ± 0.001 | 77.20 ± 3.96 |

Tab. 3 Noisy MNIST ($10^6$ samples) results and timings for different $B$ values

| B | Clustering accuracy | NMI | Execution time |
|---|---|---|---|
| Baseline | – | – | – |
| 32 | 64.19 ± 1.03 | 0.541 ± 0.005 | 2334.31 ± 25.63 |
| 64 | 60.97 ± 0.3 | 0.506 ± 0.001 | 1243.81 ± 23.43 |

### 4.5. Molecular dynamics trajectory clustering

In this section we analyze the behaviour of the clustering algorithm in terms of the quality of the obtained results in the MD domain. Basically, we compared the results obtained by the current implementation with respect to the results obtained in [1]. In that paper the binding process of a drug to its target was simulated and we used an in house clustering tool to get intermediate states of the protein/ligand complex formation along the binding routes. There, we employed the k-medoids algorithm and we were able to completely characterize the binding process.

Here we ran again the same kind of analysis systematically verifying that the same, or very similar, binding intermediates could be obtained. For the analysis of the structures, we extracted the medoids from each cluster. The same atoms as per [1] were used for the clustering. To define the number of clusters we used the elbow criterion as in [1] trying the clustering in the (4,40) range; in the end, we obtained 20 clusters as an optimal value.

For each run we initialized 5 times the algorithm with the k-means++ method and kept the solution with minimum cost. To assess the accuracy of the approximated algorithm we split the dataset in 4 mini-batches each comprising about 250000 samples, thus drastically limiting the kernel matrix size with respect to a full run. We used the strided sampling because data was batch available and when possible, this sampling should be always used. As previously anticipated, we evaluated the quality of the results by the capability of the solution to capture the key events of the simulations. In Fig.7(a) we summarize the meaning of the medoids in structural terms using the same naming conventions appeared in [1] and associate them with the respective cluster id.

Overall those medoids well recapitulate the binding process giving the same synthetic description obtained in [1] despite the mini-batch approximation. In particular, we show here, in Fig.7(b), the distance matrix computed across the medoids; we reordered the columns based on the manual classification induced by visual inspection. Results show clearly the three main macro-sections of the simulation namely the bound state, the entrance paths and the out unbound states.

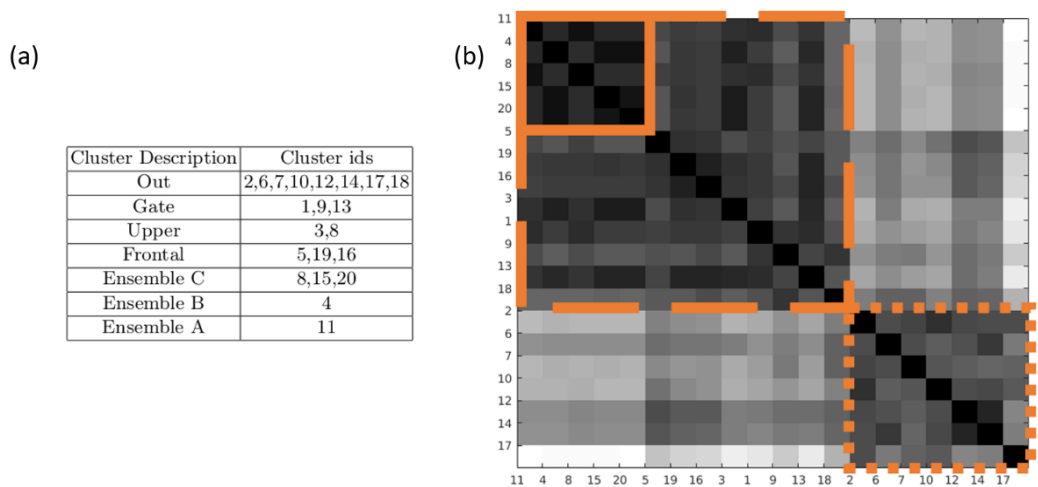

Fig. 7 (a) Table summarizing medoids for MD data and their labelling (b) Medoids RMSD matrix. On the axes we indicate the medoid identifiers. On the upper left is well visible the macro-area of the bound states. Then, this area extends to the right including the entrance paths, and lastly, on the lower right corner, the unbound states

## 5. DISCUSSION

Mini-batch approaches are not new in the clustering community and encountered a great success when applied to standard k-means [9]. In his work, Sculley showed how a mini-batch Stochastic Gradient Descent (SGD) procedure converges faster than regular GD. He proposed to set the size of mini-batches to a rather small value, namely $\approx 10^3$, and to fix an a-priori number of iterations for the algorithm.

Our take here is quite different. The number of iterations is by construction equal to the number of mini-batches $B$ in order to exploit the entire dataset. Moreover, a major difference with the SGD procedure proposed by Sculley is here represented by the inner loop. We actually believe that iterating each mini-batch up to convergence can lead to a better minimization of the cost function and to a less noisy procedure.

A comparison about the clustering accuracy achieved by the two algorithms for the original MNIST dataset is shown in Fig.8. It is worth noting that our proposed algorithm performs better as the number of mini-batches $B$ decreases whereas the performances of the SGD procedure proposed by Sculley are almost constant. Moreover, and as expected, our algorithm is less sensitive to noise, indeed the clustering accuracy variance is much lower in comparison to that of the SGC procedure.

We stress also the fact that our parallelization approach is rather different when compared to what in literature is referred to as parallel patch clustering, see e.g. [16]. Indeed, we don't parallelize across mini-batches assigning one mini-batch per node. Instead, we parallelize the iterations within each mini-batch thus allowing the algorithm to cope with virtually any sample size.

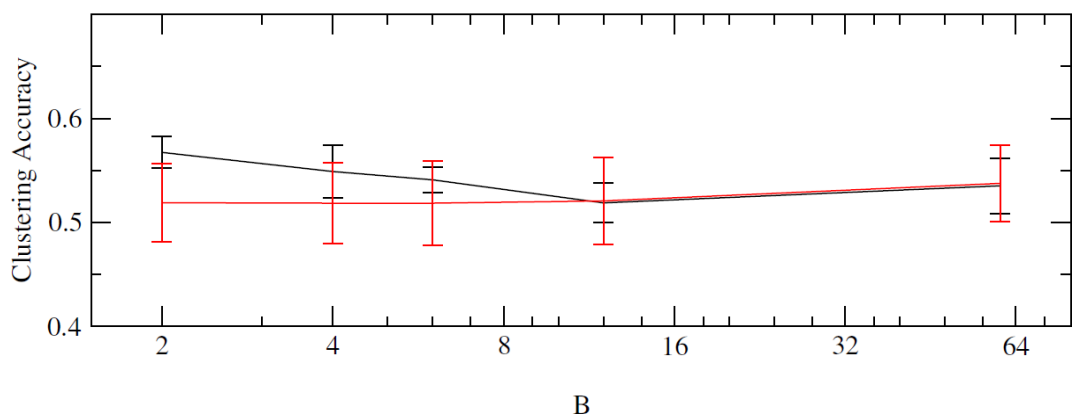

Fig. 8 Clustering Accuracy vs number of mini-batchs $B$ for the proposed algorithm (black line) and the SGD k-means procedure proposed by Sculley (red line). Comparison performed on the original MNIST dataset with $C = 10$, $\sigma = 4d_{max}$ to mimic a linear behaviour.

## 6. CONCLUSIONS

In this paper we presented a distributed and efficient approximation scheme for the kernel k-means algorithm. The approximation scheme applies an adaptive strategy based on the available memory resources together with the full exploitation of CPUs and GPUs capabilities. We obtained state of the art results in several application domains in terms of accuracy even in a heavily approximated regime; moreover, we got linear scaling in several different, distributed, computational architectures, something particularly useful in the big data era.

Next developments will deal with the full GPU porting of the algorithm exploiting GPU direct communications facilities of nVidia GPUs and the systematic application to the molecular dynamics domain, with particular attention to drug discovery, possibly proposing algorithmic extensions to best fit the field requirements.

## REFERENCES

[1] Sergio Decherchi et al., (2015), "The ligand binding mechanism to purine nucleoside phosphorylase elucidated via molecular dynamics and machine learning.", *Nature communications*, 6.

[2] Mark Girolami, (2002), "Mercer kernel-based clustering in feature space.", *IEEE Transactions on Neural Networks*, 13, 3, pp780–784.

[3] Radha Chitta, et al., (2011), "Approximate kernel k-means: Solution to large scale kernel clustering", *Proceedings of the 17th ACM SIGKDD international conference on Knowledge discovery and data mining*, ACM, pp 895–903.

[4] Luca Mollica et al., (2015), "Kinetics of protein-ligand unbinding via smoothed potential molecular dynamics simulations.", *Scientific Reports*, 5.

[5] S Kashif Sadiq et al., (2012), "Kinetic characterization of the critical step in HIV-1 protease maturation.", *Proceedings of the National Academy of Sciences*, 109, 50, pp 20449–20454.

[6] Rong Zhang and Alexander I Rudnicky. (2002), "A large scale clustering scheme for kernel k-means.", *Pattern Recognition. Proceedings. 16th International Conference on*, 4, pp289–292.

[7] Leon Bottou and Yoshua Bengio, (1995), "Convergence properties of the k-means algorithms." *Advances in neural information processing systems*, pp 585–592.

[8] David Arthur and Sergei Vassilvitskii, (2007), "k-means++: The advantages of careful seeding.", *Proceedings of the eighteenth annual ACM-SIAM symposium on Discrete algorithms. Society for Industrial and Applied Mathematics*, pp 1027–1035.

[9] David Sculley, (2010), "Web-scale k-means clustering.", *Proceedings of the 19th international conference on World wide web*, pp 1177–1178.

[10] Jason Sanders and Edward Kandrot, (2010), "CUDA by example: an introduction to general-purpose GPU programming.", Addison-Wesley Professional.

[11] Yann LeCun and Corinna Cortes, (1998), "The MNIST database of handwritten digits.".

[12] David D Lewis et al., (2004), "Rcv1: A new benchmark collection for text categorization research." *Journal of machine learning research*, 5, pp 361–397.

[13] Wen-Yen Chen et al., (2011). "Parallel spectral clustering in distributed systems.", *IEEE transactions on pattern analysis and machine intelligence*, 33, 3, pp 568–586.

[14] Meng-Chiao Ho, et al., (2010), "Four generations of transition-state analogues for human purine nucleoside phosphorylase." *Proceedings of the National Academy of Sciences*, 107, 11, pp 4805–4812.

[15] Fabian Pedregosa, et al., (2011), "Scikit-learn: Machine learning in Python. Journal of Machine Learning Research", 12, pp 2825–2830.

[16] Alex, N.and Hammer, B., (2008), "Parallelizing single patch pass clustering", *ESANN,* pp. 227-232.

## AUTHORS

**Marco Jacopo Ferrarotti**

Graduated in Physics of Complex Systems in 2013 jointly from Politecnico di Torino and Paris-Sud University. Since 2014 he moved to the Drug Discovery and Development Department of the Italian Institute of Technology as PhD student working on study and developments of scalable Machine Learning methods applied to Molecular Dynamics simulations.

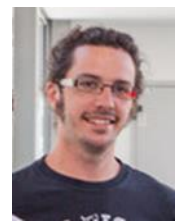

**Sergio Decherchi**

Graduated in Electronic Engineering in 2007 from Genoa University, Italy. Since 2005 he started collaborating with the Department of Biophysical and Electronics Engineering of Genoa University, where he completed a PhD in Machine Learning and Data Mining in 2010. His main research interests are computational byophisics and computational intelligence tools for drug discovery and virtual screening. He published more than 20 papers in refereed conferences and journals.

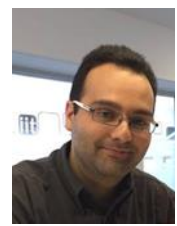

**Walter Rocchia**

Graduated in Electronic Engineering on July 1996. In February 2000, he got a PhD in Electronic Devices at the University of Trento. He then was a Research Scholar at the Biochemistry Department of the Columbia University. In 2008, he moved to the Drug Discovery and Development Department of the Italian Institute of Technology, working on computational approaches to ligand-protein binding free energy estimation. In late 2014, he created the Computational mOdelling of NanosCalE and bioPhysical sysTems (CONCEPT) Lab. He is author of more than 50 publications including International Journals, book contributions and Proceedings.

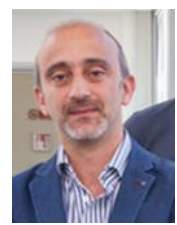